\documentclass[prl,twocolumn,floatfix,showpacs,preprintnumbers,amsmath,amssymb]{revtex4}

\usepackage[dvips]{graphics}

\usepackage{graphicx}

\begin{document}

\title{Determining the current polarization in Al/Co nanostructured point contacts}

\author{F. P\'erez-Willard,$^1$ J.C.~Cuevas,$^2$ C. S\"urgers,$^1$ P. Pfundstein,$^3$ 
J.~Kopu,$^2$ M. Eschrig,$^2$ and H. v. L\"ohneysen$^{1,4}$} 

\affiliation{$^1$Physikalisches Institut, Universit\"at Karlsruhe, 
D-76128 Karlsruhe, Germany\\
$^2$Institut f\"ur Theoretische Festk\"orperphysik, Universit\"at Karlsruhe, 
D-76128 Karlsruhe, Germany\\
$^3$Laboratorium f\"ur Elektronenmikroskopie, Universit\"at 
Karlsruhe, D-76128 Karlsruhe, Germany\\
$^4$Forschungszentrum Karlsruhe, Institut f\"ur Festk\"orperphysik, 
D-76021 Karlsruhe, Germany}

\date{\today}

\begin{abstract}
We present a study of the Andreev reflections in superconductor/ferromagnet
nanostructured point contacts. The experimental data are analyzed in the frame 
of a model with two spin-dependent transmission coefficients for the majority
and minority charge carriers in the ferromagnet. This model consistently describes 
the whole set of conductance measurements as a function of voltage, temperature, and 
magnetic field. The ensemble of our results shows that the degree of spin 
polarization of the current can be unambiguously determined using Andreev physics.
 
\end{abstract}


\maketitle

The field of spintronics is largely based on the ability of ferromagnetic 
materials to conduct spin-polarized currents~\cite{Prinz1995}. Thus, the 
experimental determination of the degree of current polarization has become
a key issue. Recently the analysis of Andreev reflections in 
superconductor/ferromagnet (S/F) point contacts has been used to extract
this spin polarization in a great variety of 
materials~\cite{Soulen1998,Upadhyay1998,Nadgorny2000,Ji2001a,Parker2002,Raychaudhuri2003}. 
The underlying idea is the sensitivity of the Andreev process to the spin 
of the carriers, which in a spin-polarized situation is manifested in a 
reduction of its probability~\cite{Jong1995}. The theoretical analysis of these
S/F point-contact experiments has been mainly carried out following the ideas of
the Blonder-Tinkham-Klapwijk (BTK) theory~\cite{Blonder1982}. Different generalizations
of this model to spin-polarized systems have been proposed, in which with an  
additional phenomenological parameter $P$, the spin polarization of the ferromagnet,
excellent fits to the experimental data have been
obtained~\cite{Soulen1998,Upadhyay1998,Nadgorny2000,Ji2001a,Parker2002,Raychaudhuri2003}. 
However, a microscopic justification of these models is
lacking~\cite{Mazin2001,Strijkers2001,Ji2001b}. Recently, Xia 
\emph{et al.}~\cite{Xia2002} have combined ab initio methods with the 
scattering formalism to analyze the Andreev reflection in spin-polarized 
systems. Their main conclusion is that, in spite of the success in fitting the
experiments, these modified BTK models do not correctly describe the
transport through S/F interfaces. Therefore, at this stage several basic
questions arise: what is the minimal model that describes on a microscopic footing
the Andreev reflection in spin-polarized systems? And, more importantly, can the
current polarization be experimentally determined using Andreev physics?

In this paper we address these questions both experimentally and theoretically.
We present measurements of the differential resistance of nanostructured Al/Co
point contacts as a function of voltage, temperature, and magnetic field. To 
analyze the experimental data we have developed a model based on quasiclassical
Green functions, the main ingredients of which are two transmission coefficients 
accounting for the majority and minority spin bands in the ferromagnet. We show 
that this model consistently describes the whole set of data, which unambiguously
demonstrates that the spin polarization of current in a ferromagnet can indeed be
determined employing Andreev reflection.

\begin{figure}[b]
\begin{center}
\includegraphics[width=0.95\columnwidth,clip=]{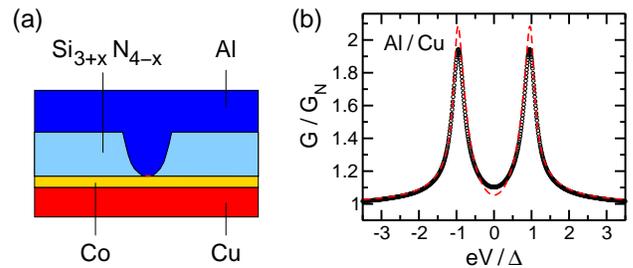}
\vspace*{-0.3truecm}
\caption{(a) Schematic of an Al/Co nanocontact. (b) Andreev spectrum of an Al/Cu
contact at 95 mK (black circles). The dashed line is the fit obtained with the BTK 
theory~\cite{Blonder1982} yielding the transmission $\tau = 0.781$ and 
the gap $\Delta= 206\,\mu$eV. }
\label{schematic_ns}
\end{center}
\end{figure}

We have fabricated Al/Co point contacts following the process described
in Ref.~\cite{Ralls1989}. Briefly, a bowl-shaped hole is drilled through a 50-nm 
thick silicon nitride (Si$_{3+x}$N$_{4-x}$) membrane by means of electron-beam 
lithography and reactive ion etching. The smallest opening in the insulating 
membrane has typically a diameter of $5\,$nm. Finally, 200~nm of Al 
and $d_{\rm Co} =6$, 12, 24 or 50 nm of Co plus ($200\,$nm - $d_{\rm Co}$) of Cu are 
deposited by electron-beam evaporation under ultra-high vacuum conditions ($\sim 
10^{-9}~$mbar) on each side of the membrane. A schematic of the samples is
shown in Fig.~\ref{schematic_ns}(a). The differential resistance $R$ was measured
with lock-in technique in a dilution refrigerator. A dc current was superimposed
to the small measuring ac component and both $R$ and the voltage drop $V$ were
recorded simultaneously. 

\begin{figure}[t]
\begin{center}
\includegraphics[width=\columnwidth,clip=]{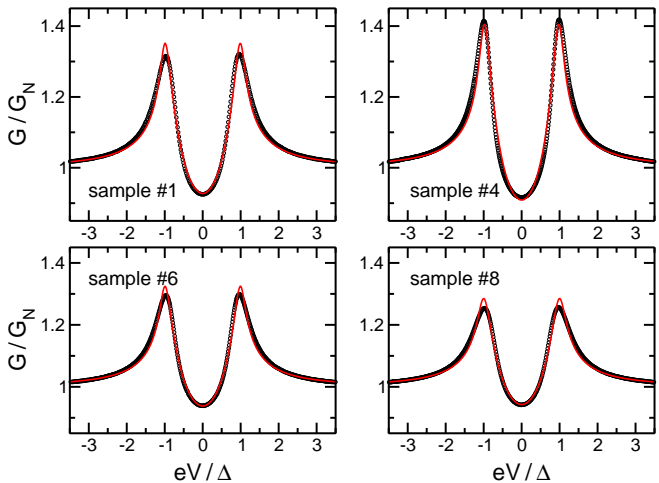}
\vspace*{-0.3truecm}
\caption{Andreev spectra of four Al/Co point contacts with different Co 
film thickness $d_{\rm Co}$. The solid line is a fit to the data with our model (see
Table~\ref{table1}).\\[-5mm]}
\label{al_co_100mK}
\end{center}
\end{figure}

As a reference we show in Fig. \ref{schematic_ns}(b) the Andreev spectrum, i.e. 
the differential conductance $G$ as a function of the voltage $V$, of an Al/Cu sample.
In all the spectra in this paper, $G$ and $V$ have been normalized by the normal state
conductance $G_N$ and by the zero-temperature superconducting gap $\Delta$ of the Al 
electrode, respectively. $G_N$ showed to be completely independent of $V$ in the range
$eV \lesssim 5-10\,\Delta$. Since the estimated mean free paths of the Cu and Al 
electrodes are $\sim 60$ nm or longer at low temperatures, all the contacts studied
are in the ballistic regime. In the Al/Cu case (Fig. \ref{schematic_ns}(b)) the BTK 
theory fits the experimental data very well (see figure caption for details). In the 
case of Al/Co, the ferromagnetic layer causes a reduction of the Andreev spectrum 
amplitude as compared to the Al/Cu contacts (see Fig.~\ref{al_co_100mK}). Notice that, 
although both the normal state resistances and the Co layer thicknesses of the
samples differ strongly (see Table~\ref{table1}), the Andreev spectra are all quite
similar. This indicates that we are observing an intrinsic property of Al/Co point
contacts.

The minimal model necessary to describe transport in S/F contacts
should account for the spin-dependent transmission, which is inherent to any junction
where ferromagnets are involved. We have developed a model that fulfills this requisite
in the framework of the quasiclassical Usadel theory~\cite{Usadel1970,Belzig1999},
describing a system in terms of two retarded Green functions, $g(\vec{r},\epsilon)$
and $f(\vec{r},\epsilon)$, which depend on both space and energy and satisfy 
$g^2 + f^2 = 1$. For transport through interfaces this theory must be
supplemented with boundary conditions, which can be formulated in terms of a normal-state
scattering matrix $\hat{S}$. Our choice to model an S/F interface is given by (we restrict 
ourselves to a single conduction channel)

\begin{eqnarray}  
\hat{S} = \left( \begin{array}{cc}
\hat{r} & \hat{t} \\
\hat{t}^{\dagger} & \hat{r}^{\prime} 
\end{array} \right) & ; &
\hat{t} = \left( \begin{array}{cc}
t_{\uparrow} & 0 \\
0 & t_{\downarrow}
\end{array} \right) ,
\hat{r} = \left( \begin{array}{cc}
r_{\uparrow} & 0 \\
0 & r_{\downarrow}
\end{array} \right) ,
\end{eqnarray}

\noindent
where $t_{\uparrow,\downarrow}$ and $r_{\uparrow,\downarrow}$ are the spin-dependent
transmission and reflection amplitudes, respectively. The transmission coefficients
$\tau_{\uparrow,\downarrow} = |t_{\uparrow,\downarrow}|^2$ are the central
quantities of our model. They contain the microscopic properties relevant for
transport, i.e. the spin-split band structure of the ferromagnet, the electronic
structure of the superconductor and the interface properties.

\begin{table}[t]
 \begin{center}
 \begin{tabular}{|c|c|c|c|c|c|c|c|c|}\hline
       sample & $d_{\rm Co}$\small{\,(nm)} & $R_N$\small{\,$(\Omega)$} & $T$\small{\,(mK)} 
		& $\Delta$\small{\,($\mu$eV)} & $\tau_\uparrow$ & 
	   $\tau_\downarrow$  & $P$ \\ \hline\hline         
       $\#1$   &  6  & 10.4 & 97 & 189 & 0.404 & 0.979 & 0.42  \\  
       $\#2$   &  6  & 6.69 & 90 & 199 & 0.403 & 0.979 & 0.42  \\
	   $\#3$   & 12  & 33.2 & 101 & 199 & 0.420 & 0.968 & 0.39  \\  
       $\#4$   & 12  & 13.3 & 100 & 188 & 0.415 & 0.970 & 0.40  \\
	   $\#5$   & 24  & 6.00 & 98 & 180 & 0.382 & 0.989 & 0.44  \\  
       $\#6$   & 24  & 3.58 & 97 & 193 & 0.399 & 0.983 & 0.42  \\
	   $\#7$   & 50  & 15.7 & 99 & 172 & 0.370 & 0.994 & 0.46  \\  
       $\#8$   & 50  & 3.59 & 97 & 198 & 0.392 & 0.986 & 0.43  \\ \hline
 \end{tabular}
 \end{center}
 \caption[]{Transmissions, $\tau_{\uparrow,\downarrow}$, polarization, $P$, and gap, 
  $\Delta$, for the Al/Co samples as determined by a fit of the Andreev spectra for 
  $T\approx 100~$mK with our model.}
  \label{table1} 
\end{table}

\begin{figure*}[t]
\includegraphics*[width=0.95\textwidth]{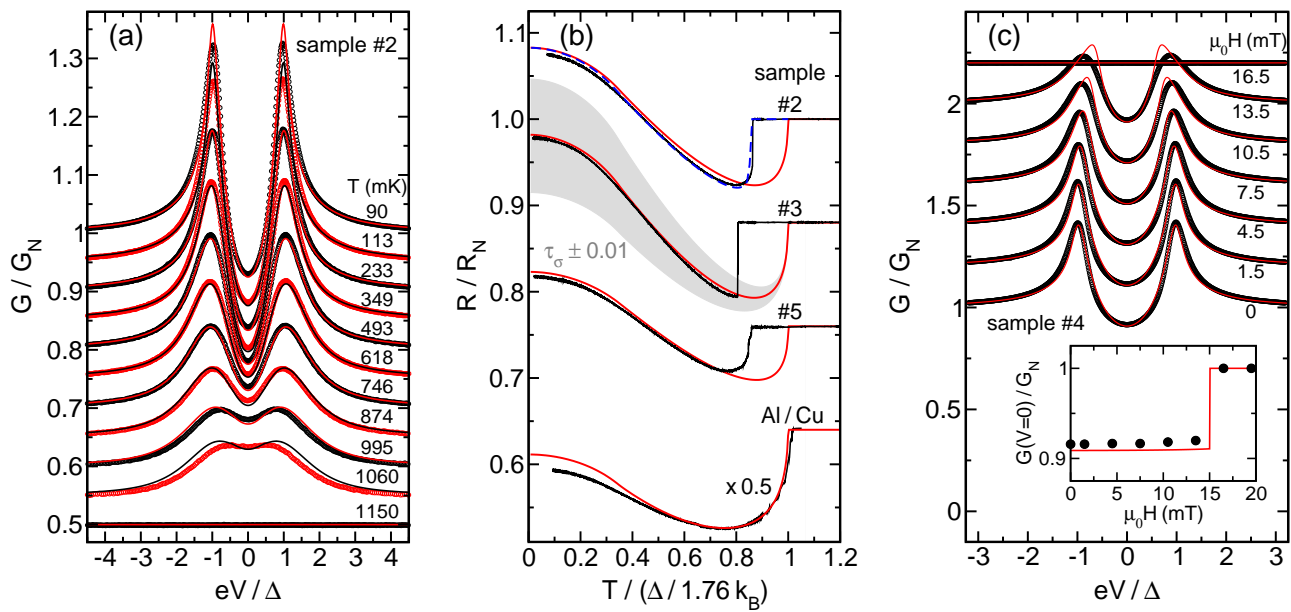}
\caption{{\bf (a)} Andreev spectrum for sample $\#2$ for different temperatures. 
For clarity, the curves are shifted downwards successively by 0.05 units with 
increasing temperature. The solid lines are the calculated spectra with our model.
{\bf (b)} Normalized resistance $R/R_N$ as a function of temperature for three Al/Co 
samples. $T$ is normalized to the gap $\Delta$ as obtained from the Andreev spectra
(see Table \ref{table1}). The curves are shifted downwards successively by 0.12 units. 
The red lines are the calculated $R(T)$. For sample $\#2$, the dashed line has been
calculated including the effect of a residual magnetic field of 5 mT. As a reference, 
we also show this data for the Al/Cu contact of Fig.~\ref{schematic_ns}(b) (the
theoretical result corresponds to the nonmagnetic BTK theory). The shaded region is
covered by a set of curves given by 
$\left\lbrace\tau_\uparrow\pm 0.01, \tau_\downarrow\pm 0.01 \right\rbrace$ 
for sample $\#3$. {\bf (c)} Andreev spectrum for sample $\#4$ measured at 100\,mK
for different magnetic fields. The curves are shifted upwards successively by 
0.2 units. The inset shows the zero-bias conductance as a function of the field. 
The critical field of the sample is $\mu_0 H_c = 15.0$ mT. The red lines are 
the calculations using $d /\lambda_0 = 3.8$.\\[-5mm]}
\label{G_dependence}
\end{figure*}

The current $I_{SF}$ through the S/F point contact is computed following 
standard procedures~\cite{Cuevas1996}. It can be separated in two spin
contributions, $I_{SF} = I_{\uparrow} + I_{\downarrow}$, where each can 
be written in the BTK form~\cite{Blonder1982}

\begin{equation}
I_{\sigma} = \frac{e}{h} \int^{\infty}_{-\infty} d\epsilon \; \left[ n_F(\epsilon-eV) -
n_F(\epsilon) \right] \left[ 1 + A_{\sigma}(\epsilon) - B_{\sigma}(\epsilon) \right],
\end{equation}

\noindent
where $n_F$ is the Fermi function, and $A_{\sigma}(\epsilon)$ and $B_{\sigma}(\epsilon)$
are the spin-dependent Andreev reflection and normal reflection probabilities, 
respectively. These are given by $A_{\sigma} = \tau_{\sigma} \tau_{-\sigma}
|f/{\cal D}|^2$ and $B_{\sigma} = |(r_{\sigma} + r_{-\sigma}) + (r_{\sigma} - r_{-\sigma})
g|^2/|{\cal D}|^2$, where $r_{\sigma} = \sqrt{1-\tau_{\sigma}}$ and ${\cal D} = (1 + 
r_{\sigma} r_{-\sigma}) + (1 - r_{\sigma} r_{-\sigma})g$. The Green functions 
are evaluated right at the interface at the superconducting side. In the point-contact
geometry we can ignore the proximity effect, which means that $g$ and $f$ only contain
properties of the superconducting electrode. In the case of a BCS superconductor in 
zero magnetic field $g=i\epsilon/\sqrt{\Delta^2 - \epsilon^2}$ and $f=-i(\Delta/
\epsilon)g$, and the zero-temperature conductance adopts the form~\cite{Cuevas2001}

\begin{equation}
G_{SF} = \frac{4e^2}{h} \left\{
\begin{array}{cc}
\frac{\tau_{\uparrow} \tau_{\downarrow}} {\left(1 + r_{\uparrow} r_{\downarrow}
\right)^2 - 4 r_{\uparrow} r_{\downarrow} (eV/\Delta)^2} & ; \; eV \le \Delta \\
\frac{\tau_{\uparrow} \tau_{\downarrow} + \left( \tau_{\uparrow} + \tau_{\downarrow}
- \tau_{\uparrow} \tau_{\downarrow} \right) \sqrt{1- (\Delta/eV)^2}}
{\left[ \left( 1 - r_{\uparrow} r_{\downarrow} \right) + \left( 1 + r_{\uparrow} 
r_{\downarrow} \right) \sqrt{1- (\Delta/eV)^2} \right]^2} & ; \; eV \ge \Delta
\end{array} \right. .
\end{equation}

\noindent
In the absence of spin polarization ($\tau_{\uparrow} = \tau_{\downarrow}$) this formula
reduces to the BTK result~\cite{Blonder1982}. The normal state conductance is given by
$G_N= (e^2/h)(\tau_{\uparrow} + \tau_{\downarrow})$, and the current polarization is 
defined by $P = |\tau_{\uparrow}-\tau_{\downarrow}|/ (\tau_{\uparrow}+\tau_{\downarrow})$. 
The main approximation of this model is the assumption that we can describe the point
contact with a single pair of transmission coefficients, $\tau_{\uparrow,\downarrow}$,
which will be finally justified by the agreement with the experiment.

As we show in Fig.~\ref{al_co_100mK}, using $\tau_{\uparrow,\downarrow}$ and $\Delta$
as free parameters our model yields an excellent fit to the Andreev spectra of the
Al/Co contacts for temperatures $T\approx 100~$mK. These parameters for a total of eight 
contacts are listed in Table \ref{table1}. Their deviations from sample to sample are
remarkably small, leading to small uncertainties in the mean
values given by $\bar{\tau}_{\uparrow}=0.40\pm0.02$, $\bar{\tau}_{\downarrow} = 
0.98\pm0.01$ and $\bar\Delta=(190\pm10)~\mu$eV. The total current
is of course symmetric with respect to the exchange of $\tau_{\uparrow}$ and
$\tau_{\downarrow}$, which implies that we cannot assign a transmission coefficient 
to the majority or minority charge carriers in Co. Nevertheless, we expect the high
transmissive coefficient $\tau_{\downarrow}$ to correspond to the minority electrons,
because of their higher density of states at the Fermi level corresponding to the 
Co 3$d$ band. In our contacts the mean value of the current polarization is
$\bar{P}=0.42\pm0.02$. An analysis of our experimental data for $T\approx 100~$mK 
with the widely used model of Ref.~\cite{Strijkers2001} gives fits of similar 
quality, but yields $\sim 15\%$ smaller values for $P$. It is important to stress 
that this model cannot be mapped onto ours, it is not rigorously founded, and 
misses the fundamental ingredient of a spin-dependent transmission.

The rest of the paper is devoted to illustrate the consistency of the model, and
in turn of the determination of the polarization $P$. We show that fixing
the set $\left\lbrace \tau_\uparrow, \tau_\downarrow\right\rbrace$ and $\Delta$,
as obtained from the spectra at $T\approx 100~$mK, the model describes without
any additional fit parameter the temperature and magnetic-field dependence 
of the conductance. For instance, in Fig.~\ref{G_dependence}(a) the 
temperature dependence of the Andreev spectrum of the sample $\#2$ is depicted.
As can be seen, the model describes the whole temperature range by simply using
the BCS temperature dependence of the gap. A more stringent test of our model is 
shown in Fig.~\ref{G_dependence}(b). Here, we compare the temperature dependence of
the zero-bias resistance with the theoretical prediction. 
The agreement is excellent, apart from the deviations close to the critical temperature. 
We attribute them to the existence of a stray field ($\sim 5\,$mT) 
created by the Co film. This idea is supported by a calculation 
(see below for details) of $R(T)$ in the presence of an external field
(Fig.~\ref{G_dependence}(b)). It is worth stressing that $R(T)$ is extremely 
sensitive to the transmission (see curve for sample $\#3$ in 
Fig.~\ref{G_dependence}(b)), which illustrates the accuracy in
the determination of $\left\lbrace \tau_\uparrow, \tau_\downarrow\right\rbrace$.

We have also measured how a magnetic field, $H$, parallel to the insulating layer modifies
the Andreev spectra (see Fig.~\ref{G_dependence}(c)). There are three main effects:
(i) the height of the two maxima diminishes with increasing field and their positions
are shifted to lower voltages. (ii) As can be seen in the inset of Fig.~\ref{G_dependence}(c),
the zero-bias conductance is constant for fields below the critical field. (iii) The 
transition to the normal state is abrupt. To understand these features we now study
how the order parameter, $\Delta$, is modified by the field. We use two approximations: 
(a) in the Al electrode the mean free path ($l \sim 60$ nm) is much smaller that the 
superconducting coherence length ($\xi_0 \sim 300$ nm), which justifies the use of the
diffusive approximation ($l \ll \xi_0$) and the Usadel theory. (b) For our Al films
$\xi_0$ is greater than the electrode thickness, $d$, which means that we can assume
that $\Delta$ and the Green functions are constant throughout the sample. With these
approximations the Usadel equation reduces to the generic equation that describes the
effect of different pair-breaking mechanisms such as magnetic impurities, supercurrents
or magnetic fields~\cite{Maki1969}

\begin{equation}
\epsilon + i \Gamma g(\epsilon,H) = i \Delta \frac{g(\epsilon,H)}{f(\epsilon,H)} 
\;\;\;\; \mbox{where} \;\;\; \Gamma = \frac{2De^2}{\hbar c^2} \langle \vec{A}^2 \rangle , 
\end{equation}

\noindent
where $D$ is the diffusion constant, $\Gamma$ is a depairing energy, which contains
the effect of the magnetic field, and $\langle \vec{A}^2 \rangle$ is the average value
of the square of the vector potential along the thickness of the Al film. Additionally, 
the order parameter $\Delta$ must be determined self-consistently~\cite{Belzig1999}.
In Al the London penetration depth is typically $\lambda_0 \sim 50$ nm, which in
our case is smaller than the thickness, $d$. This implies that the external field is
partially screened inside the sample. Thus, the vector potential appearing in Eq.~(4) 
must be determined solving the Maxwell equation: $\nabla^2 \vec{A} = -(4\pi/c) \vec{j}$,
where $\vec{j}$ is the supercurrent density given by $\vec{j}(\vec{r}) = -(2\sigma_N/\hbar c) 
\vec{A}(\vec{r}) \int^{\infty}_0 d\epsilon \; \tanh \left( \beta \epsilon/2 \right)
\mbox{Im} (f^2)$, where $\sigma_N$ is the normal conductivity of the Al sample and
$\beta = (k_B T)^{-1}$. The solution of the Maxwell equation yields the following 
expression for the depairing energy 

\begin{equation}
\Gamma(H) = \frac{6 \alpha}{r^2 \cosh^2(r/2)} \left( \frac{\sinh(r)}{r} - 1 \right)
\;\; ; \;\; \alpha = \frac{D e^2 d^2 H^2}{6\hbar c^2} ,
\end{equation}

\noindent
where $r = (d/\lambda_0) \left[ (2/\pi) \int^{\infty}_0 d\epsilon^{\prime} \; \tanh 
\left( \frac{\beta^{\prime} \epsilon^{\prime}}{2} \right) \mbox{Im} (f^2) \right]^{1/2}$.
Here, the prime indicates that the energy variables are measured in units of the
zero-temperature gap in the absence of field, $\Delta_0$, and $\lambda_0 = 
\sqrt{\hbar c^2 / (4 \pi^2 \sigma_N \Delta_0)}$. In Eq.~(5) $\alpha$ is the 
pair-breaking parameter for a thin film~\cite{Maki1969}, which can also be written as 
$\alpha / \Delta_0 = (1/12\pi) \left[H d/H_{cb} \lambda_0 \right]^2$, where $H_{cb}$ is
the bulk critical field. For Al $\mu_0 H_{cb}= 9.9$ mT. Notice that the ratio 
$d/\lambda_0$ is the only parameter that enters our analysis.
Since $d/\lambda_0$ determines the critical field of the Al films, $H_c$,
we fix its value by means of an independent measurement of $R(B)$ at 
$T\approx100\,$mK. For our samples, we find $H_c \approx 1.5H_{cb}$, which in our 
theory corresponds to $d \approx 4\lambda_0$. Thus, using Eq.~(2) with 
the self-consistent solution of Eq.~(4) for the Green functions, 
we calcute the magnetic field evolution of the Andreev spectra, reproducing 
the main experimental features without any additional parameter (see 
Fig.~\ref{G_dependence}(c)). The theoretical analysis of the critical field reveals 
that for $d > \lambda_0$, as in our case, both $\Delta$ and the spectral gap
are finite up to the transition to the normal state. This naturally explains why this 
transition is of first order and why the zero-bias conductance is not 
modified by the field. The existence of this first order transition in 
superconducting films was first discussed in the frame of the Ginzburg-Landau 
theory~\cite{deGennes1966}.

In conclusion, we have presented a comprehensive experimental study of the transport 
through Al/Co nanocontacts. We have also introduced a model for the description of 
the Andreev reflection in S/F interfaces. While retaining the simplicity of 
BTK-type theories, our model includes the effect of a spin-dependent transmission
and allows the analysis of a great variety of realistic ingredients. We have shown that
such a model consistently describes the whole set of measurements for arbitrary voltage,
temperature and magnetic field, which demonstrates that the current polarization
in ferromagnets can be determined using Andreev physics. Moreover, our data and
analysis provide important input for first principle calculations of electron
transmission through ferromagnetic interfaces.

We acknowledge the financial support provided by the Deutsche Forschungsgemeinschaft
through SFB 195 and within the Center for Functional Nanostructures.


\end{document}